\documentclass[sn-mathphys,Numbered]{sn-jnl}%

\usepackage{graphicx}%
\usepackage{multirow}%
\usepackage{amsmath,amssymb,amsfonts}%
\usepackage{amsthm}%
\usepackage{mathrsfs}%
\usepackage[title]{appendix}%
\usepackage{xcolor}%
\usepackage{textcomp}%
\usepackage{manyfoot}%
\usepackage{booktabs}%
\usepackage{algorithm}%
\usepackage{algorithmicx}%
\usepackage{algpseudocode}%
\usepackage{listings}%

\theoremstyle{thmstyleone}%

\theoremstyle{thmstyletwo}%

\theoremstyle{thmstylethree}%

\begin{document}

\title[Article Title]{Cosmic ray susceptibility of the Terahertz Intensity Mapper detector arrays}

\author*[1]{\fnm{Lun-Jun} \sur{Liu}}\email{lliu@caltech.edu}

\author[2,1]{\fnm{Reinier M.J.} \sur{Janssen}}

\author[2]{\fnm{Bruce} \sur{Bumble}}

\author[1]{\fnm{Elijah} \sur{Kane}}

\author[1]{\fnm{Logan M.} \sur{Foote}}

\author[2,1]{\fnm{Charles M.} \sur{Bradford}}

\author[1]{\fnm{Steven} \sur{Hailey-Dunsheath}}

\author[3]{\fnm{Shubh} \sur{Agrawal}}

\author[3]{\fnm{James E.} \sur{Aguirre}}

\author[4]{\fnm{Hrushi} \sur{Athreya}}

\author[3]{\fnm{Justin S.} \sur{Bracks}}

\author[4]{\fnm{Brockton S.} \sur{Brendal}}

\author[3]{\fnm{Anthony J.} \sur{Corso}}

\author[4]{\fnm{Jeffrey P.} \sur{Filippini}}

\author[4]{\fnm{Jianyang} \sur{Fu}}

\author[5]{\fnm{Christopher E.} \sur{Groppi}}

\author[5]{\fnm{Dylan} \sur{Joralmon}}

\author[6]{\fnm{Ryan P.} \sur{Keenan}}

\author[4]{\fnm{Mikolaj} \sur{Kowalik}}

\author[6]{\fnm{Ian N.} \sur{Lowe}}

\author[3]{\fnm{Alex} \sur{Manduca}}

\author[6]{\fnm{Daniel P.} \sur{Marrone}}

\author[5]{\fnm{Philip D.} \sur{Mauskopf}}

\author[6]{\fnm{Evan C.} \sur{Mayer}}

\author[4]{\fnm{Rong} \sur{Nie}}

\author[4]{\fnm{Vesal} \sur{Razavimaleki}}

\author[5]{\fnm{Talia} \sur{Saeid}}

\author[7]{\fnm{Isaac} \sur{Trumper}}

\author[4]{\fnm{Joaquin D.} \sur{Vieira}}

\affil*[1]{\orgdiv{Department of Physics}, \orgname{California Institute of Technology}, \orgaddress{\street{1200 E California Blvd}, \city{Pasadena}, \postcode{91125}, \state{CA}, \country{USA}}}

\affil[2]{\orgname{Jet Propulsion Laboratory}, \orgaddress{\city{Pasadena}, \postcode{91011}, \state{CA}, \country{USA}}}

\affil[3]{\orgname{University of Pennsylvania}, \orgaddress{\city{Philadelphia}, \postcode{19104}, \state{PA}, \country{USA}}}

\affil[4]{\orgname{University of Illinois Urbana-Champaign}, \orgaddress{\city{Urbana}, \postcode{61801}, \state{IL}, \country{USA}}}

\affil[5]{\orgname{Arizona State University}, \orgaddress{\city{Tempe}, \postcode{85287}, \state{AZ}, \country{USA}}}

\affil[6]{\orgname{University of Arizona}, \orgaddress{\city{Tucson}, \postcode{85721}, \state{AZ}, \country{USA}}}

\affil[7]{\orgname{ELE Optics}, \orgaddress{\city{Tucson}, \postcode{85705}, \state{AZ}, \country{USA}}}

\abstract{We report on the effects of cosmic ray interactions with the Kinetic Inductance Detector (KID) based focal plane array for the Terahertz Intensity Mapper (TIM). TIM is a NASA-funded balloon-borne experiment designed to probe the peak of the star formation in the Universe. It employs two spectroscopic bands, each equipped with a focal plane of four $\sim\,$900-pixel, KID-based array chips. Measurements of an 864-pixel TIM array shows 791 resonators in a 0.5$\,$GHz bandwidth. We discuss challenges with resonator calibration caused by this high multiplexing density. We robustly identify the physical positions of 788 (99.6$\,$\%) detectors using a custom LED-based identification scheme. Using this information we show that cosmic ray events occur at a rate of 2.1$\,\mathrm{events/min/cm^2}$ in our array. 66$\,$\% of the events affect a single pixel, and another 33$\,$\% affect $<\,$5 KIDs per event spread over a 0.66$\,\mathrm{cm^2}$ region (2 pixel pitches in radius). We observe a total cosmic ray dead fraction of 0.0011$\,$\%, and predict that the maximum possible in-flight dead fraction is $\sim\,$0.165$\,$\%, which demonstrates our design will be robust against these high-energy events.}


\maketitle

\section{Introduction}\label{sec1}

The cosmic star formation rate density (SFRD) peaks at redshift $\sim\,2$ ($\sim 10\,$Gyr ago) and then declines significantly ($\sim 10 \times$) to the current Universe \cite{madau2014}. Understanding the mechanism of the drastically declining cosmic SFRD since $z \sim 2$ is a compelling topic in astrophysics. During this epoch of cosmic noon, most of the active star-forming regions are heavily surrounded by dust \cite{hauser2001,lagache2005}, which is optically thick to the ultraviolet (UV) radiation emitted by newly formed stars. Measuring far-infrared (FIR) emission lines is a vital probe of the dust-obscured star-forming regions as (1) dust absorbs UV radiation and re-emits it at FIR wavelengths and (2) FIR photons do not suffer from the dust extinction. 

Motivated by the advantage of FIR wavelengths, we are designing and constructing the Terahertz Intensity Mapper (TIM), a NASA-funded balloon-borne spectrometer, to unravel the star formation and galaxy evolution during its rapid decline ($0.5<z<1.7$) \cite{marrone2022}. TIM will measure the emission of the redshifted 157.7$\,\mathrm{\mu m}$ emission line of singly ionized carbon, [CII], which is a key tracer of star formation. To perform these spectroscopic observations, TIM employs two long slit grating spectrometers covering the 240-317 $\mathrm{\mu m}$ and 317-420 $\mathrm{\mu m}$ wavelength bands with spectral resolution of $R \sim 250$ \cite{vieira2019}. Each of TIM's spectrometers is equipped with a focal plane architecture containing four quadrant arrays, each of which contains $\sim\,$900 hex-packed, horn-coupled aluminum kinetic inductance detectors (KIDs) on a single readout line. A single RFSoC-based readout system will simultaneously monitor the loading on all pixels in a single focal plane \cite{sinclair2022}.

Single pixel testing of our KID design has demonstrated a sensitivity that exceeds TIM's science requirement based on the predicted loading during flight. Specifically, (1) the detector's noise equivalent power (NEP) is $3.5 \times 10^{-18} \mathrm{W/Hz^{1/2}}$, and (2) the detector is photon noise limited at 100 fW of loading at the in-flight operation temperature of 250$\,$mK \cite{janssen2022}. The detector arrays have been fabricated and tested in a sub-scale focal plane architecture. They have demonstrated a high fabrication yield, $>\,$95$\,$\%, and a negligible susceptibility to magnetic fields of a few $\mathrm{\mu T}$, which is easily achievable with basic shielding \cite{liu2022a, liu2022b}.

Further characterization of TIM's KID-based focal plane array is required to translate the high fabrication yield into a yield of science-quality detectors. In this paper we present progress of the focal plane characterization with an emphasis on the effects of cosmic rays. In Section~\ref{sec2}, we present a brief overview of our experimental setup, including the KID array layout and our custom KID identification method, through which we identify $>\,$99$\,$\% of the resonator positions. Section~\ref{sec3} shows an improvement in our noise analysis routine, which accounts for close, but non-collided, resonances in this densely multiplexed array. Using properly calibrated noise timestreams, we quantify the effect of cosmic ray hits on the array in Section~\ref{sec4}. In Section~\ref{sec5}, we summarize our results and discuss future steps in the characterization of the TIM focal plane arrays.

\section{Experimental setup}\label{sec2}

In this paper we present measurements of an 864-pixel array of KIDs for TIM’s long-wavelength spectrometer. The lumped-element KIDs (LEKIDs), similar to those in Janssen et al. \cite{janssen2022}, consist of the same $\sim\,$500$\,\mathrm{\mu m}$ diameter inductive absorber surrounded by 3 choke rings, an interdigitated capacitor (IDC), and a coplanar capacitor coupling to the microstrip readout line. All these structures are lithographically patterned in a single layer of 30$\,$nm aluminum (Al) on top of a 6$\,$" silicon-on-insulator wafer. Centered on each absorber a 1.3$\,$mm diameter (matching the outer diameter of the largest choke ring) hole is etched through the 600$\,\mathrm{\mu m}$  carrier wafer to create a 25$\,\mathrm{\mu m}$  backshort, by depositing 100$\,$nm of Al across the entire backside.

The total array is a $90.6 \times 80.75$ mm die, of which the $36 \times 24$ pixel array covers a $85.0 \times 48.5$ mm active area in the north-east corner of the die, as shown in Fig. 1 of Liu et al. \cite{liu2022b}. To efficiently pattern the entire array with stepper lithography, we carry out a unit cell approach \cite{janssen2024}. Each unit cell contains 16 $(4 \times 4)$ LEKIDs. The resonator frequencies of these 16 KIDs are designed to be \textit{widely separated} $(df/f \sim 0.05)$ and define 16 resonator clusters within the $500 - 1000\,$MHz readout bandwidth. This is achieved by varying the number of IDC tines between 24 and 68. Stepping through each of the 54 $(9 \times 6)$ unit cells, the 16 resonances in the unit cell are designed to be shifted by $df/f \sim 7.5 \times 10^{-4}$ by reducing the IDC tine length in steps of $dL = 0.9 \, \mathrm{\mu m}$. As such, each consecutive unit cell is shifted incrementally in frequency to fill out the entire frequency bandwidth of the array.

\begin{figure}[h]%
\centering
\includegraphics[width=0.95\textwidth]{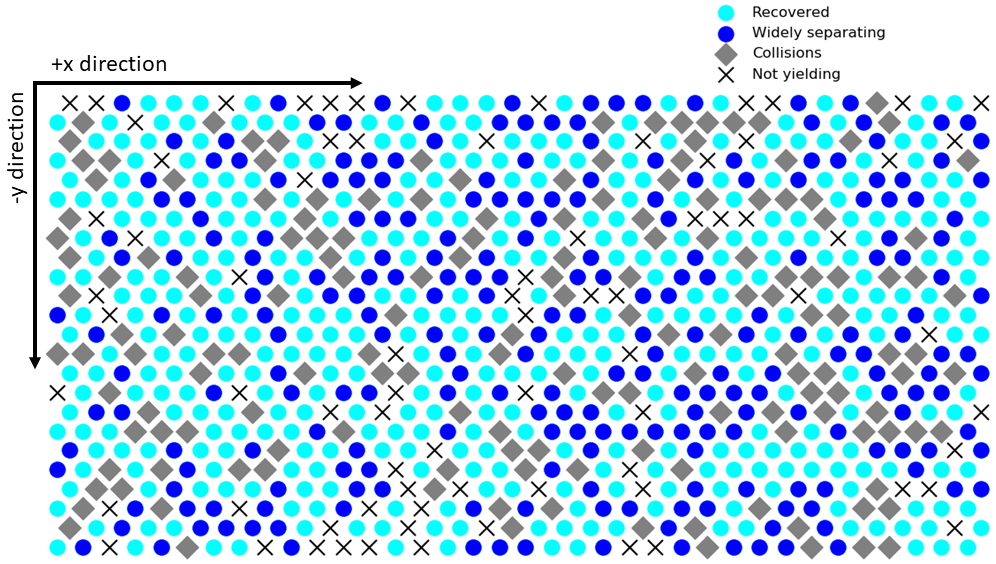}
\caption{Spatial map of TIM's 864-pixel array, with the array chip facing up. Dots and diamonds represent pixels that respond to the LED's illumination, wherein blue dots represent widely-separated, well-calibrated pixels, cyan dots have a nearby resonance feature and are properly calibrated thanks to our improved calibration scheme, and grey diamonds indicate poorly calibrated resonances, predominantly due to collisions in frequency space. Black crosses indicate resonators that did not yield.}\label{mapper}
\end{figure}

We measure the performance of this array in a pulse-tube pre-cooled triple-stage He sorption cooler at a stage temperature of 250$\,$mK. The array is located inside a sub-scale focal plane architecture similar to those designed for flight \cite{liu2022a}. Following pioneering work done by NIST \cite{liu2017a}, we use a cryogenic LED array mounted directly on top of our horn array to identify the spatial location for each of the resonance features in the readout transmission. Motivated by the concept of the unit cell design, we implement a custom designed LED mapping scheme. The key feature of this mapping method is that one single LED illuminates one unit cell (of 16 KIDs). These 16 resonators have widely separated frequencies so they remain in their designed frequency order, allowing unambiguous identification. By sequentially applying a bias to every single LED we can then provide unambiguous identification of all KIDs in the full array. Liu et al. \cite{liu2024} will provide a detailed description regarding the implementation of this custom LED KID identification mapper and its potential application to ultra-large KID arrays and small pixel pitch sizes (especially for those smaller than the LED size). Using our LED KID mapper, we robustly identify the physical location of 788 out of 791 resonators, resulting in a $>\,$99.6$\,$\% identification yield. It is worth noting that 7 of these resonances were only observed because of the LED illumination, as they were fully merged with the resonance feature of other detectors. Figure~\ref{mapper} shows the spatial map of identified resonators, plotted in blue, cyan, and grey symbols, while the black crosses are those not moved by the LED's illumination and labeled as ``not yielding.''

\section{Frequency and noise calibration in densely multiplexed KID arrays}\label{sec3}
\begin{figure}[h]%
\centering
\includegraphics[width=0.85\textwidth]{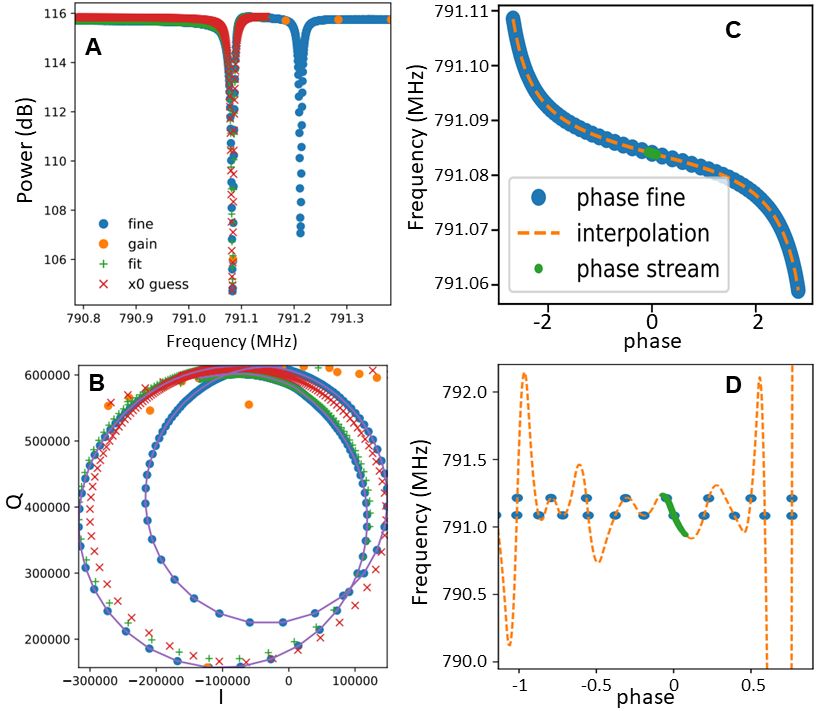}
\caption{An example showing how a resonator can be properly calibrated using our improved calibration scheme. In this example, Resonator A with $f_{res, A} = 791.084$ MHz has a neighbor, Resonator B, at $f_{res, B} = 791.212$ MHz. In \textbf{Panel A}, our improved method isolates Resonator A from Resonator B, and so Resonator A is properly fitted in green cross marks. \textbf{Panel B} demonstrates the successful fit in IQ space, enabled by our improved calibration method, while the original method tries to fit the 2 IQ loops hence failed. \textbf{Panel C} shows that the improved calibration routine enables the good fit of the frequency vs. phase diagram, whereas \textbf{Panel D} indicates the poor fit from the original calibration method, because it tries to fit the data from 2 near but isolated resonators.}\label{compare}
\end{figure}

Near neighbors and especially the collisions in the frequency space have always been regarded as a dominant source of contamination in the frequency and noise calibration process and decrease of the number of science-usable detectors. An example of their impacts on the conventional ROACH2 calibration process is demonstrated in Figure~\ref{compare}, using an example of Resonator A with a resonant frequency ($f_{res,A}$) = 791.084 MHz. In Panel A, Resonator A has a neighbor in the frequency scan range, Resonator B, with $f_{res,B} = 791.212$ MHz, and they both maintain their complete resonance features. This calibration routine is likely to be confused by the 2 partially overlapping IQ loops and the phase noise data from 2 separated resonators, so therefore this routine poorly fits the IQ loop (Panel B) and incorrectly convert the phase noise to $df/f$ noise (Panel D) given the unreasonable fit (Panel D). The high multiplexing density makes this a pronounced effect, which will also affect any future arrays pushing to higher multiplexing densities \cite{foote2023}.

The key solution to the proper calibration of the resonator frequency is to reduce the calibration frequency range centered around the nominal resonator frequency in post-processing. This ensures that the nominal resonator is properly \textit{isolated} from its near neighbors without compromising the measured frequency range required for lower-Q resonators within the array. In order for the proper fit to the data, in IQ space it is necessary to maintain an uncontaminated IQ loop within a phase range from $-\pi / 2$ to $+\pi / 2$; this phase range can be translated to the $|S_{21}|$ along the readout transmission in frequency domain --- the proper fit of a resonator feature requires at least an isolated full width at half minimum of the resonator dip. As a first demonstration, we implement an improved calibration routine that features a collision threshold in the frequency domain, $\Delta f/f = (f_{i+1}-f_{i})/f_{i} = 7 \times 10^{-5}$ ($\Delta f \sim 50\,$kHz in the middle of the readout bandwidth), which matches well to the typical $Q\sim2\times10^4$. With this threshold, every pair of 2 near resonator neighbors with spacing below this threshold will be labeled as ``collision,'' and above will be labeled as 2 isolated resonators. This improved calibration method is now able to (1) flag collisions, (2) properly separate near neighbor resonances, (3) reject drifted noise streaming data, and (4) implement robust interpolation for the conversion from phase noise data to $df/f$ noise.

\begin{figure}[h]%
\centering
\includegraphics[width=1.0\textwidth]{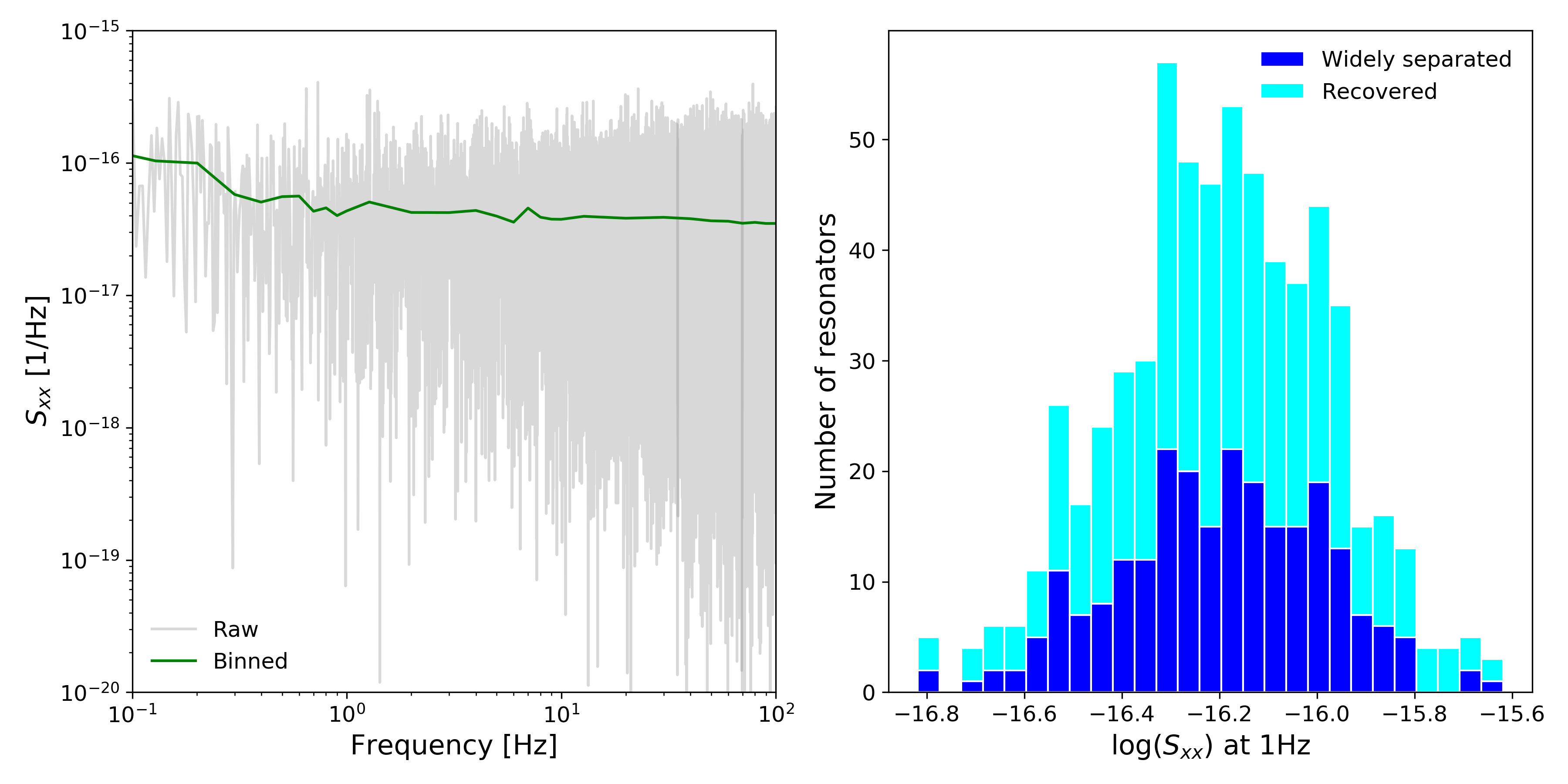}
\caption{\textbf{Left}: $S_{xx}$ of Resonator A extracted with the improved calibration routine. The grey curve shows the raw $S_{xx}$, while the green curve represents the binned $S_{xx}$ in the frequency range of interest. \textbf{Right}: The stacked distribution of $S_{xx}$ at $1\,$Hz, wherein the blue bars indicate the histogram of widely separated resonators, and the cyan bars shows the histogram of resonators recovered by our improved calibration scheme.}\label{sxx}
\end{figure}

Figure~\ref{sxx} demonstrates the success on the recovering a large number of resonators enabled by properly isolating neighbor resonators in our improved calibration scheme. Figure~\ref{sxx} (left) indicates that the recovered Resonator A's fractional frequency noise power spectral density ($S_{xx}$) is at a (white noise) level in line with others in the array --- In Figure~\ref{sxx} (right), the recovered 381 resonators follow a comparable distribution to the 248 widely separated resonators when studying their $S_{xx}$ at $1\,$Hz (the modulation frequency given TIM's fiducial scan strategy). It is noteworthy that among all properly calibrated KIDs, the average of $S_{xx}$ at $1\,$Hz is at a white noise level $\approx 6.3 \times 10^{-17} \, \mathrm{Hz^{-1}}$, comparable to our single pixel measurement (see Fig. 4 of Janssen et al. \cite{janssen2022}). This result clearly indicates that our properly calibrated KIDs indeed reach science quality. Figure~\ref{mapper} illustrates the boosted yield of proper-calibration, enabled by our improved calibration routine--the number of properly calibrated KIDs increases from 248 to 629, a great improvement of $\sim\,$150$\,$\%. Panel C of Figure~\ref{compare} describes the proper calibration enabled by our improved calibration routine, in contrast with the original poor calibration. Our improved calibration scheme is capable of isolating Resonator A from its near neighbor, Resonator B, and then properly fits both the IQ data and the phase-to-frequency conversion. As a result, we are able to recover the true $S_{xx}$ of Resonator A and all properly isolated resonators.

\section{Impact of cosmic rays on the KID array}\label{sec4}

KIDs will experience data loss when high-energy cosmic rays hit the focal plane array chip \cite{karatsu2019}. These high-energy particles hit the detector wafer and deposit energy in the substrate during TIM's balloon flight, creating athermal phonons which break Cooper pairs and generate extra quasi-particles in the KID absorbers that cause the large frequency excursion and produce spurious signals. The key to mitigation is to absorb and thermalize these hot phonons before they reach the absorbers. Techniques to achieve this have been demonstrated \cite{karatsu2019}, and have informed our design. We present our measurements, the inferred rates, and how our design mitigates the impact of cosmic rays.

We take time ordered noise streaming data and search for cosmic ray events. Using our ROACH2 data acquisition system, we simultaneously stream the noise data of all identified KIDs with sampling rate $f_{sampling} = 488\,$Hz for 5 minutes. The excursion rises within a single time interval followed by a quick decay with a time constant of $\sim\,$100 microseconds \cite{janssen2022}. We set a $5\sigma$ threshold to filter out cosmic ray event candidates, because (1) the typical non-affected resonator's $df/f$ noise falls in the $5\sigma$ range of a Gaussian distribution and (2) simulations show that the residuals of cosmic rays at levels of $<\,$5$\sigma$ in the data do not affect the RMS noise \cite{catalano2016}.

These $>\,$5$\sigma$ events allow for the calculation of the cosmic ray event rate over the entire effective area of the array chip, event rate per pixel, and the data loss fraction. We identify a total of 434 unique events (that is, events impacting multiple KIDs treated as a single event) measured by the 629-KID array in the full observation time. The area occupied by the KIDs is approximately 41.23 cm$^2$, and we thus have a lower limit\footnote{Larger fluxes could be possible if the effective area impacted by a single cosmic ray even is smaller than a pixel.} to the energetic particle flux on the chip of $2.1 \, \mathrm{events/min/cm^{2}}$. This estimate is consistent with literature, where Karatsu et al. \cite{karatsu2019} reported $\sim\,$2.5$\,\mathrm{events/min/cm^{2}}$, and Wilen et al. \cite{wilen2021} reported a combined total rate of gamma and muon particles of $\sim\,$1.9$\,\mathrm{events/min/cm^{2}}$ which was also scaled by McEwen et al. \cite{McEwen2022}.
 
We define the total data loss fraction due to cosmic ray hits as $f_{loss} = \sum\limits_{N} \sum\limits_{i} (E_{N, i}/f_{sampling}) / N_{tot} \, t_{tot}$, where $N_{tot} = 629$, $E_{N, i} = 0$ except when detector $N$ is impacted by a cosmic ray in time sample $i$, then $E_{N, i} = 1$, and $t_{tot}$ represents the 5-minute noise timestream. The $f_{loss}$ obtained in this experiment is a neglible 0.0011$\,$\%. 
 
We attribute this low $f_{loss}$ to a number of features of our architecture:  (1) Most importantly, the thick (100$\,$nm) Al deposited on the backside of the wafer and backshorts has a lower transition temperature (thus lower superconducting gap energy) compared to the 30$\,$nm Al of the absorber. As athermal phonons interact with this layer, they are down-converted, at least partially, to gap-edge excitations which are then unable to break pairs in the higher-critical-temperature KID absorbers. (2) The bulk of the cosmic ray phonons are generated in the carrier wafer, and do not have a direct path to the absorber because of the 25$\,\mu$m thick, 1.3$\,$mm diameter backshort, which provides a geometric constriction to phonons on their path from the bulk silicon to the absorber. (3) Finally, the constriction is further enhanced by the fact that the outer 0.5 mm annulus of the backshort has the thick 100 nm backshort Al on the back and 30 nm Al optical choke rings on the front. They both absorb and thermalize high energy phonons before they can reach the 500$\,\mu$m diameter absorber in the center of the backshort ``membrane.'' The first 2 features are similar to those implemented previously (e.g., Karatsu et al. \cite{karatsu2019}), and the third improvement may further reduce $f_{loss}$ compared to other literature values\footnote{This may explain our measured $f_{loss}$ $\sim\,$4$\times$ lower than the minimum value reported by Karatsu et al. \cite{karatsu2019}, though note that the sensitivity of the TIM devices at 250$\,$mK is an order of magnitude lower, which complicates the attribution.}.

\begin{figure}[h]%
\centering
\includegraphics[width=0.95\textwidth]{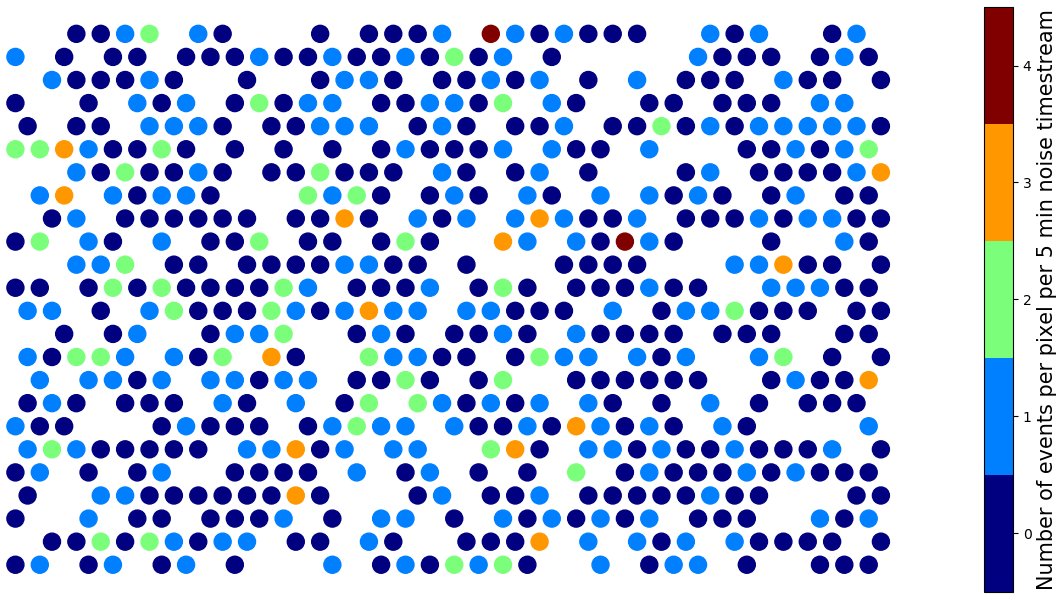}
\caption{The spatial map of events occurring on individual pixels across the entire TIM focal plane array with the color bar showing the number of cosmic ray events per pixel per 5-minute noise timestream. An animated movie of an example event is shown in the online supplementary material.}\label{cosmic_ray_frame}
\end{figure}

\begin{figure}[h]%
\centering
\includegraphics[width=0.95\textwidth]{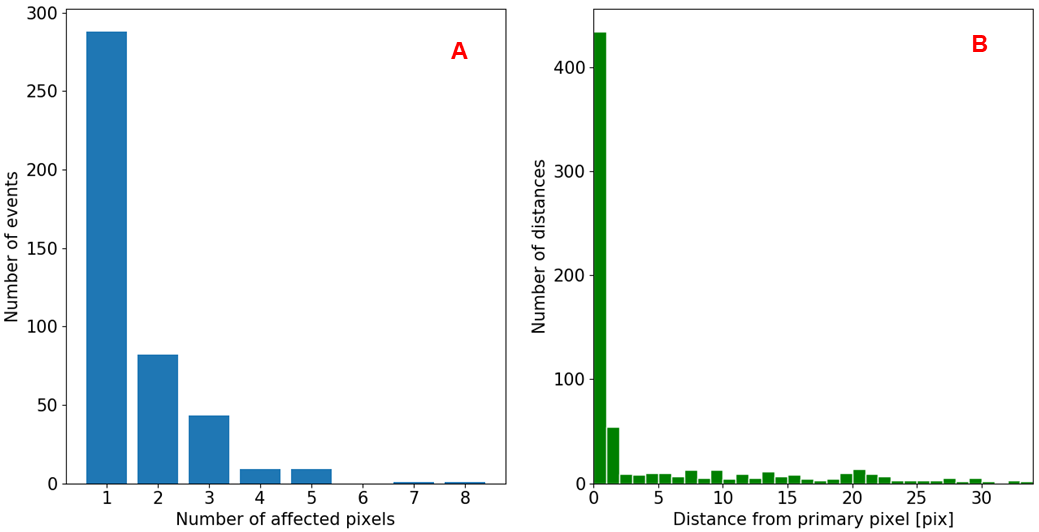}
\caption{\textbf{Panel A}: The distribution of the number of affected pixels of the cosmic ray event. \textbf{Panel B}: The distribution of distances between companion pixels and the primary pixel of time frames affected by the cosmic ray event.}\label{num}
\end{figure}

Combining the noise data and the spatial map of our KID array, we are able to further characterize the distribution of cosmic ray events in the array. Figure~\ref{cosmic_ray_frame} reports the number of cosmic ray events occurring per pixel in a 5-minute noise timestream given the individual pixel's physical location. It can be seen that the cosmic ray hits are randomly distributed across the entire wafer. The attached online supplementary shows an animation of an example cosmic ray event using snapshots of 4 consecutive frames. Figure~\ref{num} studies how localized these cosmic ray events are: Panel A indicates that $>\,$66$\,$\% of the events affect 1 single pixel, with another $33\%$ affecting 5 or fewer. Panel B indicates that the cosmic ray events are highly localized, as $>\,$60$\,$\% of the secondary noise spikes are constrained to the nearest or next-nearest neighbors of the primary pixel that first experiences the energy deposit from certain cosmic ray event. This implies a typical spread over a 0.66$\,\mathrm{cm^{2}}$ region (2 pixel pitches in radius). These results show that the cosmic ray events are highly confined, explaining in part our low $f_{loss}$.

The ground-based characterization enables the prediction of TIM KID arrays' cosmic ray susceptibility at float. Literature have reported that (1) the muon flux at float is $\sim 2-3 \,\mathrm{/sec/cm^2}$ \cite{workman2022}, and (2) the flux of energetic particles measured by \textit{Planck} space observatory is 5$\,\mathrm{/sec/cm^2}$; therefore, we upscale the lab-based cosmic ray event rate by $\sim 75 - 150$$\times$ as an estimation of the averaged cosmic ray flux in TIM's balloon flight. This leads to our predictions on the in-flight $f_{loss}$, where (1) $f_{loss, avg} \approx f_{loss, lab} \times 75 = 0.083\,$\%, and (2) $f_{loss, max} \approx f_{loss, lab} \times 150 = 0.165\,$\%. Even the upper range of this estimate is very favorable for a spaceflight detector array -- our architecture offers improvements relative to some of the previous KIDs flown, e.g., total $f_{loss} \approx 3\,$\% from the OLIMPO balloon mission's in-flight measurements with their KID arrays \citep{masi2019, Paiella2020}.

\section{Summary and prospects}\label{sec5}

We present the characterization of TIM's 864-pixel KID array in an effort to maximize the science-quality yield and investigate the cosmic ray susceptibility. We have mapped 788 of the 791 yielded KIDs to their physical location. We have also improved our frequency and noise calibration method and then applied it to these identified resonators, obtaining 629 non-collided, properly calibrated resonators, a $\sim\,$150$\,$\% improvement from the original calibration method. 

We further demonstrate that these recovered resonators show an averaged white noise level of $S_{xx}$ in line with widely separated ones and comparable to our single-pixel measurement, which clearly indicates their high science quality. We have characterized the susceptibility of KIDs to cosmic ray events. We confirm that the cosmic ray event flux in our experiment is comparable to that reported by others, and we show that our array architecture offers excellent suppression of the impact of these events on the KIDs. Events are highly localized and result in a negligible data loss fraction in the lab, and expected losses of much less than 1\% in the stratospheric balloon float or spaceflight environment. We attribute this excellent performance to a phonon thermalization layer on the back of our array and the geometry of our choke rings and the backshort.

To further increase the number of science usable detectors, we are planning to rearrange collided resonators' $f_{res}$ by trimming the capacitors folllowing the techniques demonstrated by NIST \cite{liu2017b} as well as the SuperSpec team \cite{mcgeehan2023}. We will also combine the spatial and frequency information associated with every single pixel to analyze the noise correlation coefficient between the pixels in the array. Based on the nature of these correlations, we will apply suitable mitigation strategies. Ultimately, we aim to turn all yielding KIDs into properly separated and science-usable detectors.

\backmatter

\bmhead{Supplementary information}

In the online supplementary material, we show an animation of an example cosmic ray event that records the entire series of the interaction between this particular cosmic ray hit and the behavior of surrounding affected pixels, from the generation of the highly perturbed $df/f$ noise to the decay back to equilibrium.

\bmhead{Acknowledgments}

TIM is supported by NASA under grant 80NSSC19K1242, issued through the Science Mission Directorate. Part of this research was carried out at the Jet Propulsion Laboratory, California Institute of Technology, under a contract with the National Aeronautics and Space Administration (80NM0018D0004).

\bibliography{sn-bibliography}%

\end{document}